\documentclass[]{spie}  

\usepackage{amsmath,amsfonts,amssymb}
\usepackage{graphicx}
\usepackage[colorlinks=true, allcolors=blue]{hyperref}
\usepackage{setspace}
\usepackage{soul}
\usepackage{lineno}
\newcommand{\arcsec}{$^{\prime\prime}$}

\title{Exploring the trade-space of distributed aperture telescopes for faint-object spectroscopy}

\author[*a]{Theodore A. Grosson}
\author[b,a]{Deborah M. Lokhorst}
\author[b,a]{Alan W. McConnachie}
\affil[a]{University of Victoria, 3800 Finnerty Road, Victoria, Canada}
\affil[b]{NRC Herzberg Astronomy and Astrophysics Research Centre, 5071 W Saanich Rd, Victoria, Canada}

\authorinfo{*tgrosson@uvic.ca}

\begin{document} 
\maketitle

\begin{abstract}
Scientific programs targeting the faintest objects push modern telescopes to increasingly large sizes, high costs and complex designs. Obtaining spectra of low surface brightness galaxies, for example, requires hours of observations on state-of-the art instruments. Increasing telescope diameters beyond 40 m raises potentially insurmountable challenges for design and funding. An innovative strategy for combating these costs is to use many small telescopes in place of a single large aperture. For observations in which high angular resolution is not necessary, this can be equivalent to a large diameter telescope in terms of sensitivity, while diffraction limited by the individual aperture. Improvements in commercial off-the-shelf (COTS) components have made this concept a possibility, as demonstrated by instruments such as the Dragonfly Telephoto Array, the Huntsman Telescope, and the Argus Array. In this work, we discuss the merits of ``distributed aperture telescopes'' as applied to the use case of spectroscopic observations of ultra-diffuse galaxies (UDGs). We compare the cost and simulated scientific performance of different configurations of apertures, detectors, and other components for this purpose, finding that an array of half-metre telescopes can obtain comparable observations to large telescopes for a fraction of the cost. Finally, we discuss the path to prototyping an array which will enable a spectroscopic survey of UDGs.
\end{abstract}


\section{Introduction}
\label{sec:intro}

Massive ($\gtrsim8$~m) telescopes, while at the technological forefront of astronomy, are extremely expensive, technically complex, and have high demands on their observing time. The 39~m Extremely Large Telescope (ELT), for example, has a total construction cost of nearly 1.5 billion euros, plus millions more each year in operating costs, as well as a decades-long development timeline\cite{EuropeanSouthernObservatory2023}. The faintest objects in the sky require massive light buckets like the ELT to study, making it very costly to learn about these objects using traditional large telescopes.

A novel approach aiming to reduce these costs which has seen recent success is the use of distributed aperture telescopes, such as the Dragonfly Telephoto Array\cite{Abraham2014}, the Huntsman Telescope\cite{Spitler2019}, the Argus Array\cite{Law2022}, and several others. Instead of a single large aperture, these telescopes build up a large collecting area with tens to hundreds of small apertures operated in parallel. For observations not requiring high angular resolution, coadding exposures from each aperture is optically equivalent to an exposure from a single large aperture. This concept is feasible due to advances in cheap, low-noise detectors, resulting in read noise not being a limiting factor for large numbers of coadds. By using mass-produced components, the cost of building such an array can be greatly reduced compared to custom-built massive components.

Distributed aperture telescopes have been shown to be broadly applicable to astronomy, especially for large field of view observations.\cite{Law2022,Abraham2022} Dragonfly has been especially successful in the identification of low surface brightness galaxies, including those termed ``ultra-diffuse galaxies'' (UDGs).\cite{vanDokkum2015} UDGs are highly extended, low surface brightness ($r_e>1.5~\rm{kpc},~\mu_{0,g}>24~\rm{mag~as^{-2}}$) galaxies whose existence are difficult to explain with typical galaxy formation pathways.\cite{Gannon2026a} Although UDGs are now understood to be relatively common, there remains a lack of spectroscopic data on them that could help constrain their nature.\cite{Gannon2026a}

In this Proceeding, we take UDGs as a case study to explore the benefits of distributed aperture telescopes in spectroscopy. In Section \ref{sec:scaling}, we motivate the use of arrays by describing their costs relative to the costs of massive telescopes. In Section \ref{sec:simulations}, we show how an array can be used to produce comparable target signal-to-noise as that of a state-of-the-art instrument, MUSE, and we constrain the design of such an array to be optimized for a survey of UDGs. In Section \ref{sec:prototype}, we discuss the path to prototyping such an array, and we state our conclusions in Section \ref{sec:conclusions}.

\section{Scaling between Cost and Aperture Size}
\label{sec:scaling}

A. Meinel\cite{Meinel1979} and G. van Belle\cite{vanBelle2004} have demonstrated that the cost of telescopes scales as a power law against their apertures, with telescopes using single monolithic mirrors built between 1980 and 2004 scaling as $D^{2.5}$. With this relation, telescopes become increasingly cost-inefficient with size, and any telescopes larger than $\sim$10~m become unrealistic to fund. The advent of giant segmented mirror (GSM) telescopes resulted in telescopes cheaper than this trend, but as of 2004 there were too few GSM telescopes to measure a power law for this type of telescope.

\begin{figure}[ht]
    \centering
    \includegraphics[width=0.8\linewidth]{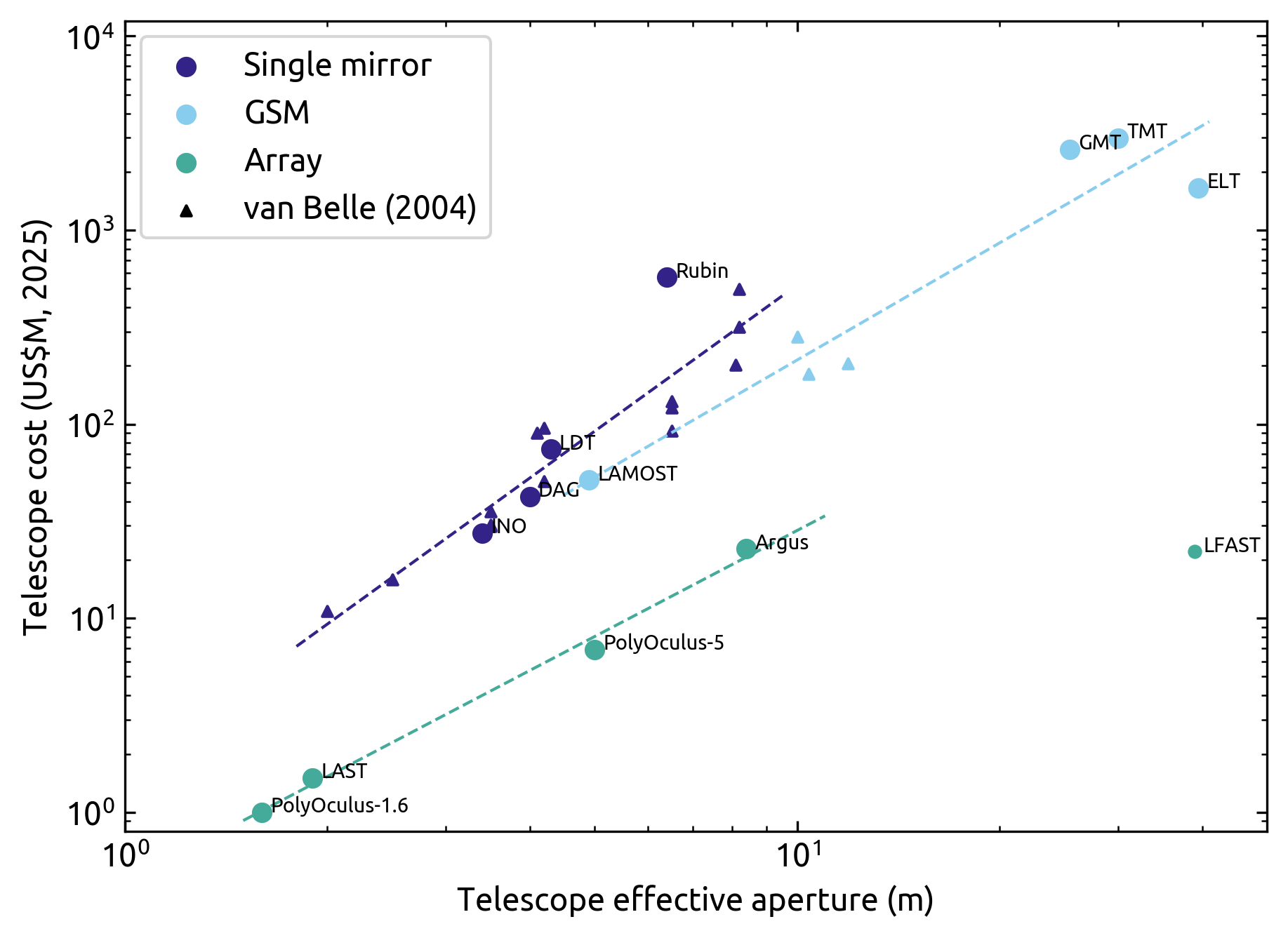}
    \caption{Cost of telescopes vs their effective aperture, along with the best-fitting power law for each type of telescope. Telescopes not included in Ref. \citenum{vanBelle2004} are labelled. LFAST has been excluded from the array fit as an outlier.}
    \label{fig:van-belle-update}
\end{figure}

We investigate how modern telescopes, including arrays, compare to the previously identified power laws by fitting power laws including additional telescopes built since 2004. As in Ref. \citenum{vanBelle2004}, we divide the telescopes between those with single monolithic mirrors and GSMs. We further add distributed aperture telescopes as a separate class, considering only reflecting telescopes for uniformity. The telescopes we include are listed in Table \ref{tab:telescopes}, as are the sources for the cost estimates that we use. Best fit power laws are shown in Figure \ref{fig:van-belle-update}. We make an effort to include only total construction cost without operating cost, and we report values in 2025 U.S. dollars using the U.S. Bureau of Labor Statistics inflation values\footnote{\url{https://data.bls.gov/timeseries/CUUR0000SA0}}.

\begin{table}[htp]
\caption{Telescopes used to fit cost trends. Costs have been normalized to 2025 U.S. dollars. Costs have been compiled from published papers as well as publicly available observatory and news websites.}
\label{tab:telescopes}
\begin{center}       
\begin{tabular}{|l|c|c|c|l|}
    \hline
    \multicolumn{1}{|c|}{Telescope} & \multicolumn{1}{|c|}{Effective} & \multicolumn{1}{|c|}{Cost} & \multicolumn{1}{|c|}{First light} & \multicolumn{1}{|c|}{Reference} \\
    & \multicolumn{1}{|c|}{aperture} &&& \\
    & \multicolumn{1}{|c|}{(m)} & \multicolumn{1}{|c|}{(US\$M, 2025)} & & \\
    \hline
    \multicolumn{5}{|c|}{Single monolithic mirror} \\ \hline
    WHT & 4.2 & 95.3 & 1987 & \citenum{vanBelle2004} \\ \hline
    NOT & 2.5 & 15.8 & 1988 & \citenum{vanBelle2004} \\ \hline
    NTT & 3.5 & 35.4 & 1989 & \citenum{vanBelle2004} \\ \hline
    ARC & 3.5 & 29.9 & 1994 & \citenum{vanBelle2004} \\ \hline
    WIYN & 3.5 & 30.4 & 1994 & \citenum{vanBelle2004} \\ \hline
    VLT (single UT) & 8.2 & 498.8 & 1998 & \citenum{vanBelle2004} \\ \hline
    Gemini (single telescope) & 8.1 & 201.9 & 1999 & \citenum{vanBelle2004} \\ \hline
    Subaru & 8.2 & 317.8 & 1999 & \citenum{vanBelle2004} \\ \hline
    MMT & 6.5 & 92.4 & 2000 & \citenum{vanBelle2004} \\ \hline
    Magellan 1 & 6.5 & 121.5 & 2000 & \citenum{vanBelle2004} \\ \hline
    Magellan 2 & 6.5 & 130.9 & 2002 & \citenum{vanBelle2004} \\ \hline
    SOAR & 4.2 & 50.9 & 2002 & \citenum{vanBelle2004} \\ \hline
    Faulkes & 2 & 10.9 & 2004 & \citenum{vanBelle2004} \\ \hline
    VISTA & 4.1 & 90.1 & 2009 & \citenum{vanBelle2004} \\ \hline
    Lowell Discovery Telescope (LDT) & 4.3 & 74.3 & 2012 & \textsuperscript{a} \\ \hline
    Iranian National Observatory (INO) & 3.4 & 27.5 & 2022 & \textsuperscript{b} \\ \hline
    Vera C. Rubin Observatory & 6.4 & 571 & 2025 & \textsuperscript{c} \\ \hline
    Eastern Anatolia Observatory (DAG) & 4 & 42.3 & 2025 & \textsuperscript{d} \\ \hline
    \multicolumn{5}{|c|}{Giant segmented mirror} \\ \hline
    LBT & 11.9 & 205.7 & 2005 & \citenum{vanBelle2004} \\ \hline
    Giant Magellan Telescope (GMT) & 25.4 & 2606.8 & 2030s & \textsuperscript{e} \\ \hline
    Keck (single telescope) & 10 & 282.7 & 1993 & \citenum{vanBelle2004} \\ \hline
    GTC & 10.4 & 181.7 & 2006 & \citenum{vanBelle2004} \\ \hline
    Large Sky Area Multi-Object Fiber & 4.9 & 51.6 & 2008 & \textsuperscript{f} \\ 
    Spectroscopic Telescope (LAMOST) &&&& \\ \hline
    Extremely Large Telescope (ELT) & 39.5 & 1648.3 & 2029 & \citenum{EuropeanSouthernObservatory2023} \\ \hline
    Thirty Meter Telescope (TMT) & 30 & 2985.4 & 2030 & \textsuperscript{g} \\ \hline
    \multicolumn{5}{|c|}{Array} \\ \hline
    Argus Array & 8.4 & 22.8 & 2027 & \citenum{Law2022} \\ \hline
    Large Array Survey Telescope (LAST) & 1.9 & 1.5 & 2023 & \citenum{Ofek2023} \\ \hline
    Large Fiber Array Spectroscopic & 39 & 22 & 2020s & \citenum{Angel2022} \\ 
    Telescope (LFAST) &&&& \\ \hline
    PolyOculus-1.6 & 1.6 & 1 & 2020s & \citenum{Eikenberry2019} \\ \hline
    PolyOculus-5 & 5 & 6.9 & 2020s & \citenum{Eikenberry2019} \\ \hline
\end{tabular}
\end{center}
{\scriptsize\singlespacing\textsuperscript{a}\href{https://web.archive.org/web/20260211120638/https://lowell.edu/research/telescopes-and-facilities/ldt/}{https://lowell.edu/research/telescopes-and-facilities/ldt}/\\
\textsuperscript{b}\href{https://web.archive.org/web/20221020030012/https://www.science.org/content/article/door-open-iranian-astronomers-seek-collaborations-their-new-world-class-telescope}{https://www.science.org/content/article/door-open-iranian-astronomers-seek-collaborations-their-new-world-class-telescope}\\
\textsuperscript{c}\href{https://spacenews.com/first-rubin-observatory-images-released-amid-concerns-about-budget-cuts}{https://spacenews.com/first-rubin-observatory-images-released-amid-concerns-about-budget-cuts}/\\
\textsuperscript{d}\href{https://web.archive.org/web/20230201020758/https://www.science.org/content/article/we-put-everything-it-modest-telescope-could-have-big-impact-turkish-science}{https://www.science.org/content/article/we-put-everything-it-modest-telescope-could-have-big-impact-turkish-science}\\
\textsuperscript{e}\href{https://web.archive.org/web/20260216051548/https://www.giantmagellan.org/2024/03/01/national-science-board-announces-federal-investment-recommendation/}{https://giantmagellan.org/2024/03/01/national-science-board-announces-federal-investment-recommendation}/\\
\textsuperscript{f}\href{https://web.archive.org/web/20251109124532/https://english.cas.cn/newsroom/archive/china_archive/cn2009/200909/t20090923_43400.shtml}{https://english.cas.cn/newsroom/archive/china\_archive/cn2009/200909/t20090923\_43400.shtml}\\
\textsuperscript{g}\href{https://web.archive.org/web/20260528171301/https://www.independent.co.uk/news/world/americas/hawaii-telescope-cost-mauna-kea-a9409756.html}{https://www.independent.co.uk/news/world/americas/hawaii-telescope-cost-mauna-kea-a9409756.html}\\
All websites accessed 5 Feb 2026.}
\end{table} 

We confirm previous results for monolithic mirrors, finding that cost~$\propto D^{2.5\pm0.2}$. However, unlike Ref. \citenum{vanBelle2004}, who expected GSM telescopes to follow the same slope at a lower cost, we find that GSMs have a cost $\propto D^{2.0\pm0.3}$. That is, GSM telescopes do not become less cost-effective per collecting area at larger sizes. As might be expected, the power laws for monolithic and GSM telescopes overlap when the total telescope size is on the order of a single mirror segment. For observations requiring high angular resolution, GSMs still offer a cost-effective telescope format.

For low--spatial resolution requirements though, arrays offer a much cheaper alternative. Excluding the Large Fiber Array Spectroscopic Telescope (LFAST), whose design attempts to minimize price as a guiding principle using custom components,\cite{Angel2022} we find that arrays have a cost~$\propto D^{1.81\pm0.1}$---arrays become cheaper per area at larger sizes. This is likely an effect of economies of scale, as arrays tend to contain highly replicatable parts as opposed to requiring single custom-built components. For the same effective aperture, a telescope array can be an order of magnitude cheaper than a GSM telescope, indicating that pursuing arrays as an alternative telescope design is a compelling strategy.

\section{Optimizing a Telescope Array for Spectroscopic Observations of Ultra-Diffuse Galaxies}
\label{sec:simulations}

We aim to demonstrate how distributed aperture telescopes can be applied to spectroscopy with the example of ultra-diffuse galaxies. In order to better understand the dynamics and composition of UDGs, it is necessary to obtain high-quality spectral observations of them. Spectra are critical for determining the distances and dynamical masses of galaxies, and they can also be used to probe their stellar content and history. UDGs are especially difficult to obtain spectra of due to their highly extended nature, typically several arcseconds across. However, spatially-resolved information of the galaxies is also highly useful for constraining different formation scenarios.\cite{Gannon2026a}

\subsection{Replicating MUSE Observations of an Ultra-Diffuse Galaxy}
\label{sec:df2}
Useful observations can be conducted by integrating the light across large areas of an integral field unit such as the VLT Multi-Unit Spectroscopic Explorer (MUSE). Refs. \citenum{Emsellem2019} and \citenum{Fensch2019}, for example, use 5~hours of integrated-light MUSE observations to identify the dynamical mass and stellar populations of the UDG NGC~1052-DF2. They also obtain spatially-resolved dynamical information by binning the integrated light into several regions. Due to the wealth of data provided by this observation, we use it as a baseline against which we can measure the performance of a hypothetical telescope array whose goal is to measure spectra of UDGs.

We simulate the capabilities of such an array using a custom exposure time calculator implementing various commercial off-the-shelf (COTS) components. The telescope body can consist of several sizes of PlaneWave\footnote{https://planewave.com/} and ASA\footnote{https://www.astrosysteme.com/} telescope and mount systems. We use a Shelyak LhiresIII\footnote{https://www.shelyak.com/produit/spectroscope-lhires-iii/?lang=en} spectrograph as our disperser, as it allows easy mounting on a telescope and resolution capabilities similar to that of MUSE. We use various CCD and CMOS detectors, though we almost exclusively use a Sony IMX455 in this study due to its low price and low noise relative to other sensors. In order to allow spatially-integrated observations, we also use a Thorlabs 7-fiber round-to-linear fiber bundle\footnote{https://www.thorlabs.com/round-to-linear-fiber-optic-bundles} as an input feed to the spectrograph. The exposure time calculator tracks the cost and throughput of all input components and returns the total cost of the telescope along with the simulated observations.

\begin{figure}[ht]
    \centering
    \includegraphics[width=0.8\linewidth]{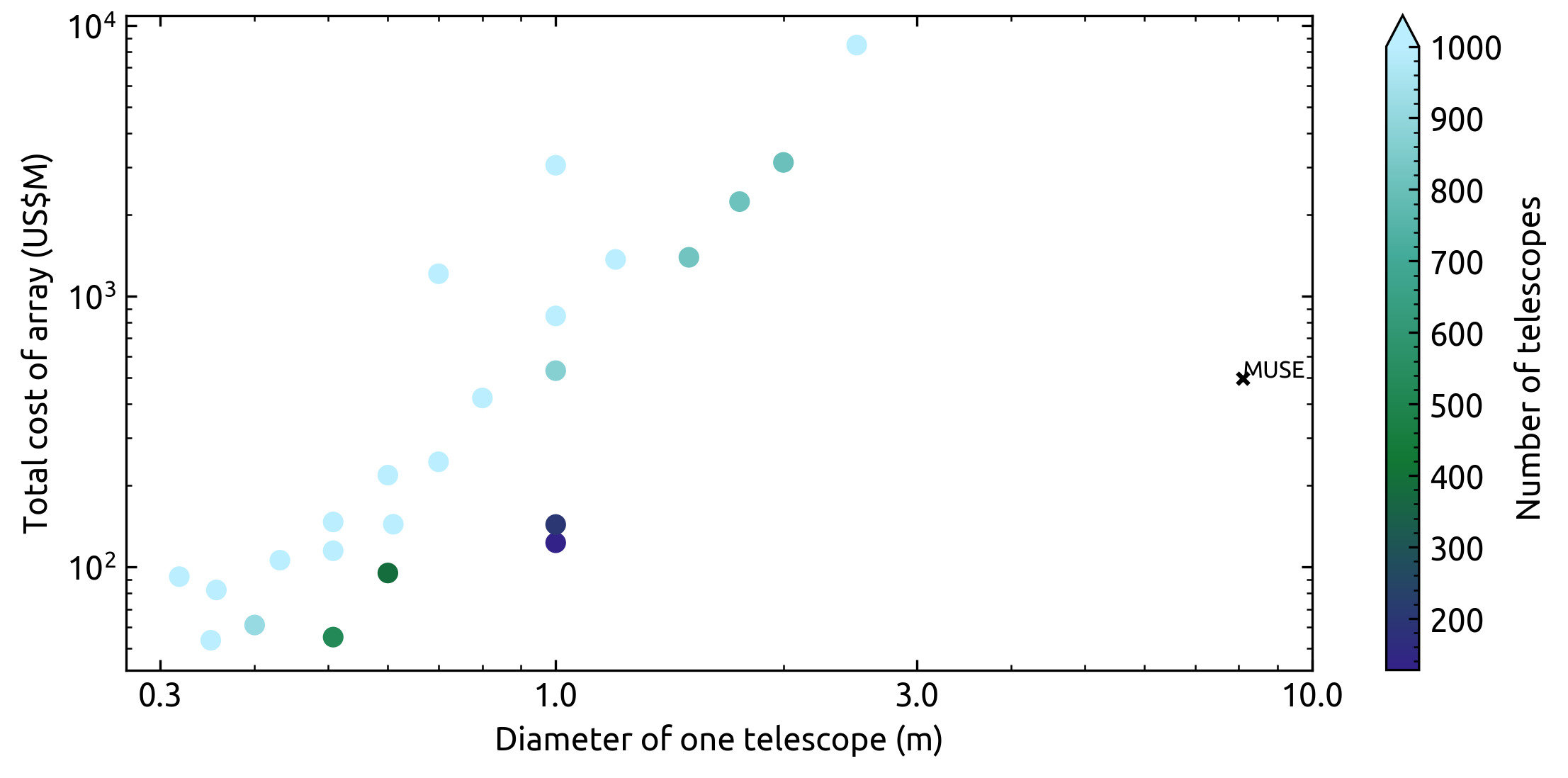}
    \caption{Cost to replicate observations from Ref. \citenum{Emsellem2019}---i.e., achieve $S/N=72.8$ in 5 hours for NGC 1052-DF2---using different COTS telescopes. Points are coloured by the number of telescopes it would take to achieve this.}
    \label{fig:DF2-cost-vs-aperture}
\end{figure}

We use a template spectrum from the eMiles library\cite{Vazdekis2016,Rock2016} using the stellar populations identified in Ref. \citenum{Fensch2019} and the photometric and structural properties from Ref. \citenum{vanDokkum2018}. After adding the median atmospheric background at Mauna Kea ($V=20.3~\rm{mag~as^{-2}}$) and noise contributions from the detector, we calculate the number of telescopes necessary, and thus the total cost, to achieve the same signal-to-noise ratio as Ref. \citenum{Emsellem2019} in the same amount of time: $S/N=72.8$ in 5~hours. The cost for various COTS telescopes as a function of their diameter is shown in Figure \ref{fig:DF2-cost-vs-aperture}. The cost of VLT + MUSE is also shown for reference.

Not all the telescopes perform well, with some arrays costing more than a VLT, but we find that replicating a MUSE observation of DF2 for a lower cost is achievable. With existing COTS components, a 0.5~m wide-field telescope performs best, using 500 telescopes for \$60M. In this case, widefield telescopes perform best due to the large size of NGC1052-DF2 ($r_e=22.6$\arcsec{}). As will be shown later, these telescopes perform more poorly for more typical UDG sizes.

For the purposes of studying UDGs, it is not necessary to use the same inputs as Ref. \citenum{Emsellem2019}. It would be beneficial to carry out a uniform spectroscopic survey of UDGs representative of the full population, so we instead use the average $r_e$ taken from the SMUDGes catalogue of UDGs, 8\arcsec{}.\cite{Zaritsky2023} Additionally, in order to measure velocity dispersions precise enough to characterize dynamical masses of these systems, it is only necessary to obtain $S/N \approx 15$. We therefore use as a base case a $S/N=15$ observation of an 8\arcsec{} scaled-down version of DF2, with the same $\mu_0$ and ellipticity. We use long exposure times to minimize read noise, since we do not expect any detector saturation for these observations, and we limit exposure times to 1~hour to account for varying sky brightness over the course of a night.\cite{Patat2008}

\begin{figure}[b]
    \centering
    \includegraphics[width=0.8\linewidth]{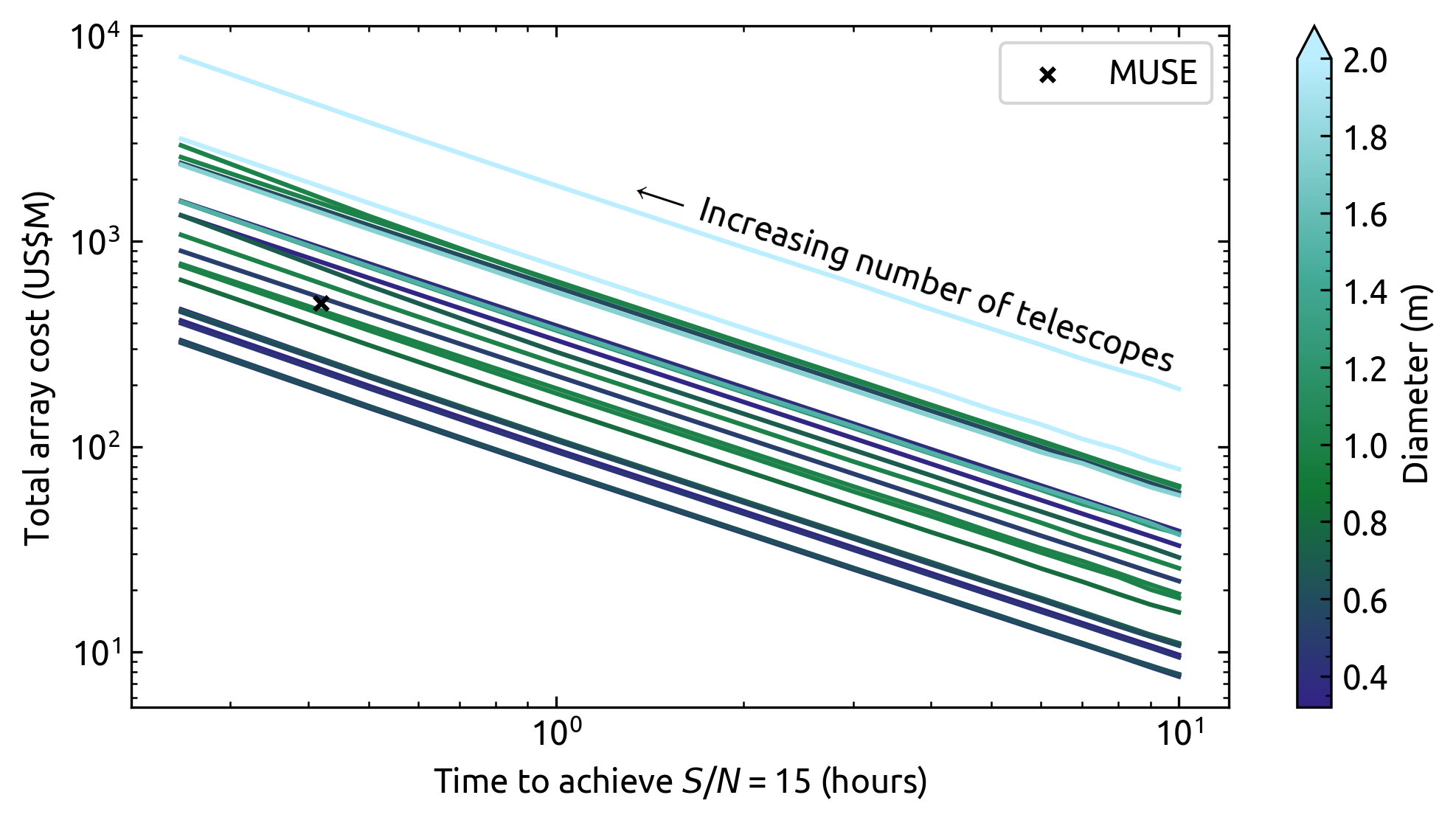}
    \caption{Tradeoff between cost and observation time for different telescope sizes, for an 8\arcsec{} UDG. Half-metre-class telescopes tend to result in the most cost-efficient observations.}
    \label{fig:cost-vs-time}
\end{figure}

It is also not necessary to complete these observations in only 5~hours, since hypothetically there would be no competition for observing time on a custom-built array. Since the full array is composed of identical unit telescopes, increasing the number of telescopes is equivalent to increasing the total observing time by the same factor, assuming approximately equal observing conditions. Therefore, cost can be lowered further by increasing the amount of observing time spent on each target. The trade-off for each of the telescopes from before is shown in Figure \ref{fig:cost-vs-time}. For each telescope, the ideal location along the cost--time curve depends on the budget and time constraints of the scientific program. \$10M worth of half-metre telescopes ($\sim$100 units, equivalent to a 5~m aperture), for example, can achieve $S/N = 15$ in a single night for an 8\arcsec{} UDG.

\subsection{An Array-Based Survey of Ultra-Diffuse Galaxies}
\label{sec:survey}

A dedicated telescope array would be capable of conducting a spectroscopic survey of UDGs. To identify the ideal components for such a survey, we arbitrarily set a construction budget of \$10M, and calculate the time it would take to achieve $S/N=15$ for a random sample of 100 galaxies from the SMUDGes catalogue. Due to the extremely diffuse nature of these objects, there are some galaxies which take a prohibitively long time to observe, with half the survey time being dedicated to only 12 galaxies of the random sample. We therefore discard galaxies which take longer than 100 hours to observe with an array of 0.5~m telescopes and resample back to 100 galaxies.

\begin{figure}[ht]
    \centering
    \includegraphics[width=0.8\linewidth]{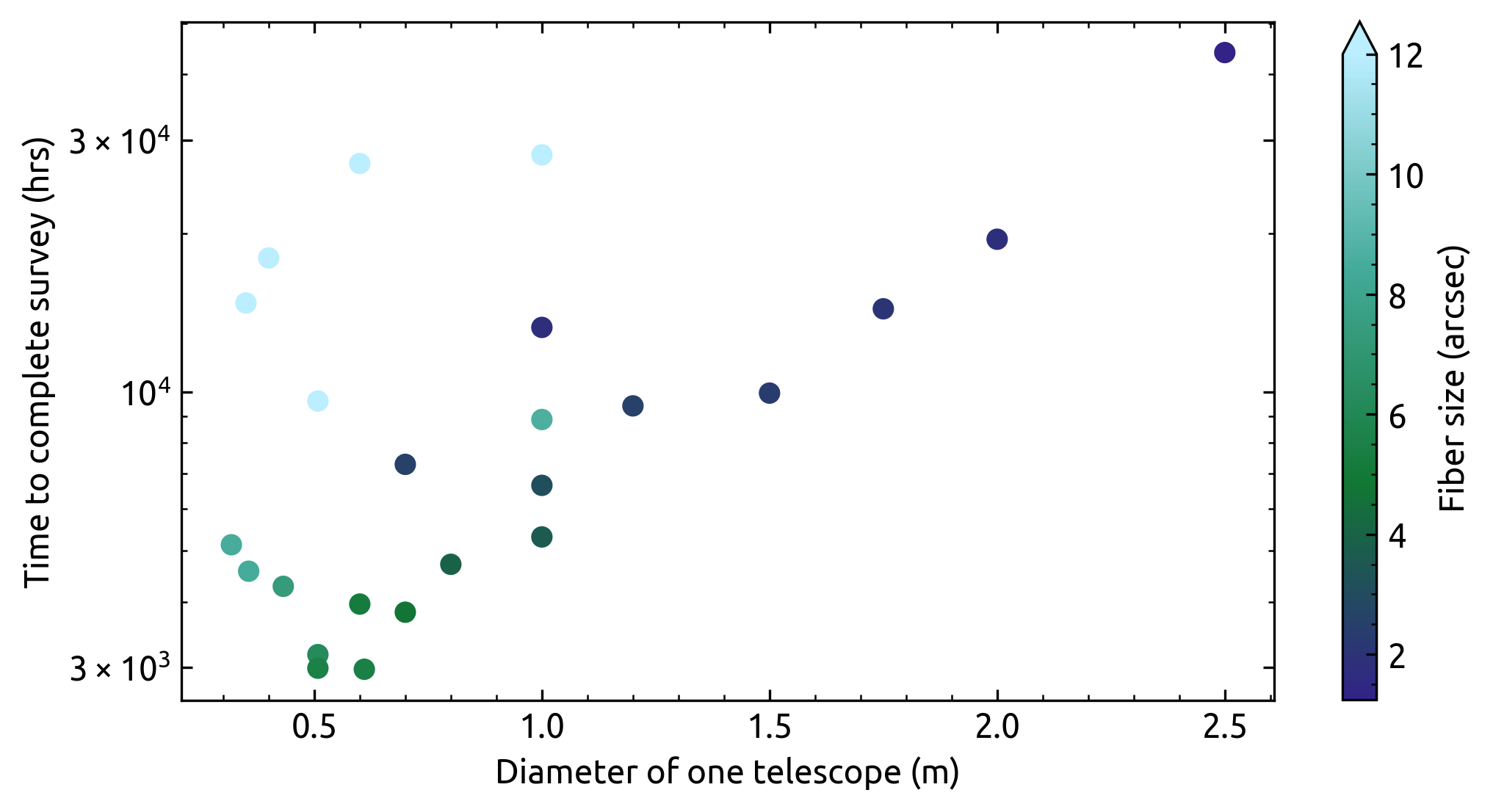}
    \caption{Time to complete a survey of 100 UDGs using a telescope array with a construction budget of \$10M. Points are coloured by the projected fiber size associated with each telescope.}
    \label{fig:survey-time}
\end{figure}

In Figure \ref{fig:survey-time}, we show the time it would take to complete this 100-UDG survey with an array consisting of \$10M worth of telescopes for each COTS telescope. We once again find that telescopes around 0.5~m perform best, with the PlaneWave CDK600 and CDK550 telescopes completing the survey in 3,000 hours, or about one year assuming an average of 8 observing hours per night. The figure also shows that the best telescopes are those with projected fiber sizes well-matched to the typical sizes of UDGs. The optimal telescopes have focal lengths resulting in projected fiber sizes slightly smaller than the average UDG $r_e$, with survey time increasing quickly for both smaller and larger fiber sizes.

\section{PROTOTYPING A TELESCOPE ARRAY}
\label{sec:prototype}

With this understanding of the ideal telescope size, we proceed to discuss the path toward prototyping an individual telescope which would constitute the array. Thus far, we have only considered the ideal cases of the components we are interested in using, including interfaces between them all. Design and construction of a prototype telescope will require some thought on how best to integrate all the individual components.

We plan to acquire a PlaneWave CDK550 telescope, one of the two most cost-efficient telescopes simulated, testing it at the Dominion Astrophysical Observatory in Victoria, Canada. We have already acquired a Shelyak Lhires III spectrograph with a 600 line/mm grating and Thorlabs 7-fiber round-to-linear fiber bundle, and we are currently in the process of characterizing them. We expect to use a Sony IMX455 sensor, though we have not yet finalized this decision.

Our simulations have not accounted for optical imperfections in these components. Although we do not require very high image quality due to our large projected fiber sizes, stray light and unexpected throughput losses could reduce the efficiency of observations. Our prototype will need to thoroughly characterize any losses relative to the idealized case here. Additional effort will need to go into efficiently coupling light between the components, especially the fiber bundle.

An additional challenge lies with the detector pixel sampling. Commercial development has driven CMOS detectors to very low noise properties, which is ideal for astronomical observations, but it has also led to small pixels. Small pixels are a benefit for high spatial resolution imaging, but in our case they result in a significantly oversampled spectrum, such that even the low per-pixel read noise becomes a dominant noise source in observations. We mitigate this by reimaging the spectrum onto the detector such that it is Nyquist sampled, with a conservatively-estimated 50\% throughput loss associated with the reimaging optics. The reimaging optics may require a custom solution which loses the advantage of COTS pricing, but it is necessary in the absence of low-cost, low-noise detectors with larger pixels.

\section{Conclusions}
\label{sec:conclusions}

Design and prototyping of this telescope will be conducted over the next couple years. The successes and failures of the process will determine the necessary components and feasibility of constructing a full array capable of carrying out a spectroscopic survey of UDGs.

We believe that existing components are capable of completing a low-cost survey in a reasonable amount of time. We simulate than an array of 100 0.5~m telescopes could survey 100 UDGs to $S/N=15$ in about one year, althought this idealized system does not account for the full range of observing conditions that can be expected. The construction cost of this array would nominally be \$10M, although we expect that, for an array of 100 units, economies of scale would provide further cost savings. This array would additionally be easy to scale to larger or smaller sizes to meet budget or survey time constraints. For any array size, we find that $\sim$0.5~m is the most cost-efficient telescope size for obtaining spatially-integrated spectra of UDGs.

Even if our specific concept proves infeasible due to design or construction challenges, we emphasize that distributed aperture telescopes are a proven method of conducting high quality observations at low cost. As we have shown, arrays offer cost savings of an order of magnitude relative to equivalent-area large telescopes. They also do not suffer from diminishing returns with increasing area; they in fact become more cost-efficient at larger sizes. This places arrays as a strong alternative to massive telescopes for science requiring apertures on the scale of 30~m and beyond.

 

\bibliography{main}
\bibliographystyle{spiebib}

\end{document}